\documentclass[runningheads,a4paper]{llncs}

\usepackage{amsfonts}
\usepackage{dsfont}

\usepackage{pifont}
\newcommand{\cmark}{\ding{51}}%
\usepackage{mathtools}

\usepackage{amssymb}

\usepackage{amsmath}

\usepackage{amsthm}

\usepackage{graphicx}

\usepackage{epstopdf}

\usepackage[]{hyperref}

\usepackage{caption}

\usepackage{array,longtable}

\usepackage{float}

\usepackage{subcaption}

\makeatletter

\def\holdocspecials{\do\ \do\$\do\&%
  \do\#\do\^\do\^^K\do\_\do\^^A\do\%}

\def\holtt{\trivlist \item[]\if@minipage\else\vskip\parskip\fi
\leftskip\@totalleftmargin\rightskip\z@
\parindent\z@\parfillskip\@flushglue\parskip\z@
\@tempswafalse \def\par{\if@tempswa\hbox{}\fi\@tempswatrue\@@par}
\obeylines \tt \let\do\@makeother \holdocspecials
 \frenchspacing\@vobeyspaces}

\makeatother

\newlength{\hsbw}
\newcommand\HOLSpacing{13pt}

   \newcommand\hilbert{\varepsilon}

   \newcommand{\Cond}{\(\Longrightarrow\)}
   \newcommand{\Eqv}{\(\equiv\)}
   \newcommand{\Iff}{\(\Leftrightarrow\)}
   \newcommand{\Fa}{\(\scriptsize \forall\)}
   \newcommand{\Et}{\(\exists\)}
   \newcommand{\Eu}{\(\exists_{unique}\)}

   \newcommand{\Func}{\(\to\)}
   \renewcommand{\Bar}{\(\mid\)}

   \newcommand{\Lam}{\(\lambda\)}
   \newcommand{\Plus}{\(+\)}
   \newcommand{\Minus}{\(-\)}
   \newcommand{\Prime}{\('\)}
   \newcommand{\Und}{\_}
   \newcommand{\Lt}{\(<\)}
   \newcommand{\Gt}{\(>\)}
   \newcommand{\Leq}{\(\leq\)}
   \newcommand{\Geq}{\(\geq\)}
   \newcommand{\Eq}{\(=\)}

\newcommand{\Hilbert}{\(\hilbert\)}
\newcommand{\Imp}{\(\Rightarrow\)}
\newcommand{\Conj}{\(\wedge\)}
\newcommand{\Disj}{\(\vee\)}
\newcommand{\Neg}{\(\neg\)}
\newcommand{\Pnd}{\(\Diamond\)}

\long\def\rechol#1#2#3{\let\next=\rechol\def\postnext{#2#3}\ifx#1\end
\let\next=\relax\def\postnext{\relax}
\else\ifx#1!\Fa                                          
\else\ifx#1@\Hilbert                                     
\else\ifx#1\#\Pnd                                        
\else\ifx#1'\Prime                                       
\else\ifx#1~\Neg                                         
\else\ifx#1\~\Neg
\else\ifx#1_\Und                                         
\else\ifx#1+\Plus
\else\ifx#1\/\Disj                                       
\else\ifx#1\.\Lam                                        
\else\ifx#1>\ifx#2=\Geq\def\postnext{#3}\else\Gt\fi      
\else\ifx#1?\ifx#2!\Eu\def\postnext{#3}\else\Et\fi       
\else\ifx#1-\ifx#2\>\Func\def\postnext{#3}               
	    \else\Minus\fi				 
\else\ifx#1|\ifx#2-\Turns\def\postnext{#3}\else\Bar\fi   
\else\ifx#1<\ifx#2=\ifx#3>\Iff\def\postnext{}            
                   \else\Leq\def\postnext{#3}\fi         
            \else\Lt\fi
\else\ifx#1=\ifx#2=\ifx#3>\Imp\def\postnext{}            
                   \else\Eqv\def\postnext{#3}\fi         
            \else\ifx#2>\Cond\def\postnext{#3}
                 \else\Eq\fi\fi
\else\ifx#1/\ifx#2\^^M\Conj\par\def\postnext{#3}         
            \else\ifx#2\ \Conj\ \def\postnext{#3}\else#1\fi\fi  
\else#1\fi\fi\fi\fi\fi\fi\fi\fi\fi\fi\fi\fi\fi\fi\fi\fi\fi\fi
\expandafter\next\postnext}

\usepackage{epsfig}

\usepackage{mdframed}

\usepackage{multirow}

\usepackage{url}

\PassOptionsToPackage{hyphens}{url}\usepackage{hyperref}

\urldef{\mailsa}\path|{adnan.rashid, osman.hasan}@seecs.nust.edu.pk|

\newcommand{\keywords}[1]{\par\addvspace\baselineskip
\noindent\keywordname\enspace\ignorespaces#1}

\begin{document}

\mainmatter  

\title{Formalization of Transform Methods using HOL Light}

\titlerunning{Formalization of Transform Methods}

%
%

\author{Adnan Rashid \and Osman Hasan}

%
\authorrunning{A. Rashid and O. Hasan}


\institute{School of Electrical Engineering and Computer Science (SEECS)\\
National University of Sciences and Technology (NUST)\\
Islamabad, Pakistan\\
\mailsa\\
\url{http://save.seecs.nust.edu.pk/projects/tm.html}
}

%
%
\maketitle

\begin{abstract}

Transform methods, like Laplace and Fourier, are frequently used for analyzing the dynamical behaviour of engineering and physical systems, based on their transfer function, and frequency response or the solutions of their corresponding differential equations.
In this paper, we present an ongoing project, which focuses on the higher-order logic formalization of transform methods using HOL Light theorem prover.
In particular, we present the motivation of the formalization, which is followed by the related work.
Next, we present the task completed so far while highlighting some of the challenges faced during the formalization.
Finally, we present a roadmap to achieve our objectives, the current status and the future goals for this project.

\end{abstract}

\section{Introduction} \label{SEC:Intro}

Differential equations are indispensable for modeling the dynamical behaviour of continuous-time engineering and physical systems.
Transform methods, which include the Laplace and Fourier transform, have been widely used for the differential equation based dynamical analysis of these systems.
These transform methods are the integral based methods, which convert a time varying function into its corresponding $s$ or $\omega$-domain representations based on Laplace and Fourier transform, respectively.
Moreover, this transformation converts the differential and integral operators in the time domain to their corresponding algebraic operators, namely, multiplication and division, in Laplace ($s$) or Frequency ($\omega$) domain and thus the arithmetic manipulation of the resulting equations involving these operators becomes easier.
These equivalent representations of the differential equations can further be used for the transfer function and frequency response analysis of these continuous-time systems.
Laplace transform is used for the analysis of the systems with causal input, whereas, in the case of non-causal input systems, Fourier transform is used.
This analysis varies in its complexity depending on the size, design parameters, constraints and the nature of the input and output signals.
The Laplace and Fourier transform methods have been widely used for the analysis of many continuous-time systems, as shown in Table~\ref{TAB:applications_of_transform_methods}.


\begin{footnotesize}
    \begin{longtable}{|p{2cm}|p{3cm}|p{7cm}|}
\caption{Applications of Transform Methods} \vspace{0.2cm}
\label{TAB:applications_of_transform_methods}
\endfirsthead
\endhead
    \hline
    \hline
    \multicolumn{1}{l}{Laplace Transform}   &
    \multicolumn{1}{l}{\hspace{0.0cm} Fourier Transform}

     \\ \hline \hline


   \multicolumn{1}{l}{{$\begin{array} {lcl} \textrm{ \hspace{-0.2cm} Control Systems~\cite{ogata1970modern,nise2007control,dorf1998modern,fortmann1977introduction,bogart1982laplace}   } \\
\textrm{ \hspace{-0.2cm} Analog Circuits~\cite{thomas2016analysis} \hspace{2.5cm} }  \\
\textrm{ \hspace{-0.2cm} Power Electronics~\cite{rashid2009power,abad2016power} }  \\
\textrm{ \hspace{-0.2cm} Astronomy~\cite{beerends2003fourier,boyce1969elementary,hilbe2012astrostatistical} }  \\
\textrm{ \hspace{-0.2cm} Mechanical Systems~\cite{oppenheim1996signals,bogart1982laplace} } \\
\textrm{ \hspace{-0.2cm} Nuclear Physics~\cite{mclachlan2014laplace,stacey2007nuclear} }
 \end{array}$}} &

   \multicolumn{1}{l}{{$\begin{array} {lcl} \textrm{  \hspace{-0.1cm} Analog Circuits~\cite{thomas2016analysis,siebert1986circuits} }  \\
\textrm{ \hspace{-0.1cm} Signal Processing~\cite{papoulis1977signal,gaydecki2004foundations,devasahayam2012signals,chu2008discrete} } \\
\textrm{ \hspace{-0.1cm} Image Processing~\cite{dougherty2009digital} \hspace{2.7cm}  }  \\
\textrm{ \hspace{-0.1cm} Biomedical Imaging~\cite{devasahayam2012signals} }  \\
\textrm{ \hspace{-0.1cm} Communication systems~\cite{ziemer2006principles,du2010wireless,madhow2014introduction,chapin1978communication} }  \\
\textrm{ \hspace{-0.1cm} Mechanical Systems~\cite{oppenheim1996signals} } \\
\textrm{ \hspace{-0.1cm} Optics~\cite{gaskill1978linear,stark2012application} } \\
\textrm{ \hspace{-0.1cm} Electromegnatics~\cite{davidson2005computational,jancewicz1990trivector,kriezis1992electromagnetics}  }
 \end{array}$}}


    \end{longtable}

\end{footnotesize}


Traditionally, the transform methods based analysis is done using paper-and-pencil proof and computer simulation methods, such as symbolic and numerical methods. However, due to the human-error proneness of paper-and-pencil proof methods and the presence of unverified symbolic algorithms, discretization errors and numerical errors in the simulations methods, the accuracy of the analysis cannot be ascertained. This in turn can lead to compromising performance and efficiency of the underlying system.  Interactive theorem proving~\cite{hasan2015formal} allows us to overcome these limitations by providing support for logic-based modeling of the system and its intended behaviour and verifying their relationship based on deductive reasoning within the sound core of a theorem prover.
With the same motivation, the Laplace~\cite{taqdees2013formalization} and Fourier~\cite{rashid2016formalization} transforms  have been formalized in higher-order logic.
In the paper, we mainly describe the past, ongoing and the planned activities for this project\footnote{\url{http://save.seecs.nust.edu.pk/projects/tm.html}}, which was started in System Analysis and Verification (SAVe) lab\footnote{\url{http://save.seecs.nust.edu.pk}} in 2012.
The formalization of Laplace~\cite{taqdees2013formalization} and Fourier~\cite{rashid2016formalization} transforms has been developed using the multivariate calculus theories of HOL Light~\cite{hollight2017multivariate}.
These formalizations also include the formal verification of some of the classical properties, such as, existence, linearity, frequency shifting, modulation, time reversal, differentiation and integration in time-domain.
We choose HOL Light theorem prover for the transform methods based analysis due to the presence of the multivariate calculus theories~\cite{hollight2017multivariate}, which contain an extensive reasoning support for differential, integral, transcendental and topology theories.

The rest of the paper is organised as follows: Section~\ref{SEC:related_work} presents the related work. The proposed approach for the transform methods based analysis is presented in Section~\ref{SEC:fremwork_for_trans_meth_analysis}.
We present the mathematical and formal definitions of transform methods, some of their formally verified classical properties and their mutual relationship in Section~\ref{SEC:formal_defs_of_transform_methods}. Section~\ref{SEC:discussion} provides the detail about the tasks that have been completed so far, the challenges faced during the formalization of transform methods, the current status and the future goals in this project. We present some of the case studies to illustrate the usefulness of the formal transform based analysis in Section~\ref{SEC:case_studies}. Finally, Section~\ref{SEC:Conclusion} concludes the paper.

\section{Related Work} \label{SEC:related_work}

Fast Fourier transform (FFT) is used for the computation of discrete Fourier transform (DFT), which is used for the analysis of the systems with the discrete-time input. Theorem provers, such as ACL2, HOL and PVS have been used for the verification of different FFT algorithms. Gamboa~\cite{gamboa1998mechanically,gamboa2002correctness} mechanically verified the correctness of FFT using a simple proof of FFT proposed by Misra using the ACL2 theorem prover. This proof utilizes the powerlist data structures, which enable the modeling of FFT using recursive functions in an efficient way and can handle the verification of many complex FFT algorithms.
Similarly, Capretta~\cite{capretta2001certifying} formalized the FFT and inverse Fourier transform (iFT) in Coq. The author used structural recursion to formalize the FFT. Whereas, the iFT is formalized using a different data type to facilitate formal reasoning about the summation operation. Moreover, isomorphism is used to link both of these data types.
Similarly, Akbarpour et al.~\cite{akbarpour2004methodology} used the HOL theorem prover for the formal specification and verification of a generic FFT algorithm. The authors used real, complex, IEEE floating point and fixed-point arithmetic theories of HOL to perform the error analysis of the FFT algorithms at real, floating-point and fixed-point levels.

Harrison~\cite{harrison15fourier} formalized the Fourier series for a real-valued function in the HOL Light theorem prover. The formalization includes the formal definition of Fourier series and formal verification of some of its properties. Similarly, Chau et al.~\cite{chau2015fourier} formalized the Fourier coefficient formulas and their properties in ACL2(r). Fourier series and their formalizations presented in HOL Light and ACL2(r) can only cater for the systems with inputs represented as periodic functions. Recently, Z-transform~\cite{siddique2014formalization} has also been formalized in the HOL Light theorem prover. However, Z-transform can only be utilized for the analysis of the systems with discrete-time input functions and cannot cater for the continuous-time systems, which is the main focus of the current paper.

\section{Proposed Approach}\label{SEC:fremwork_for_trans_meth_analysis}

\begin{figure}[!t]
\centering
\scalebox{0.27}
{\includegraphics[trim={5.0 0.4cm 5.0 0.4cm},clip]{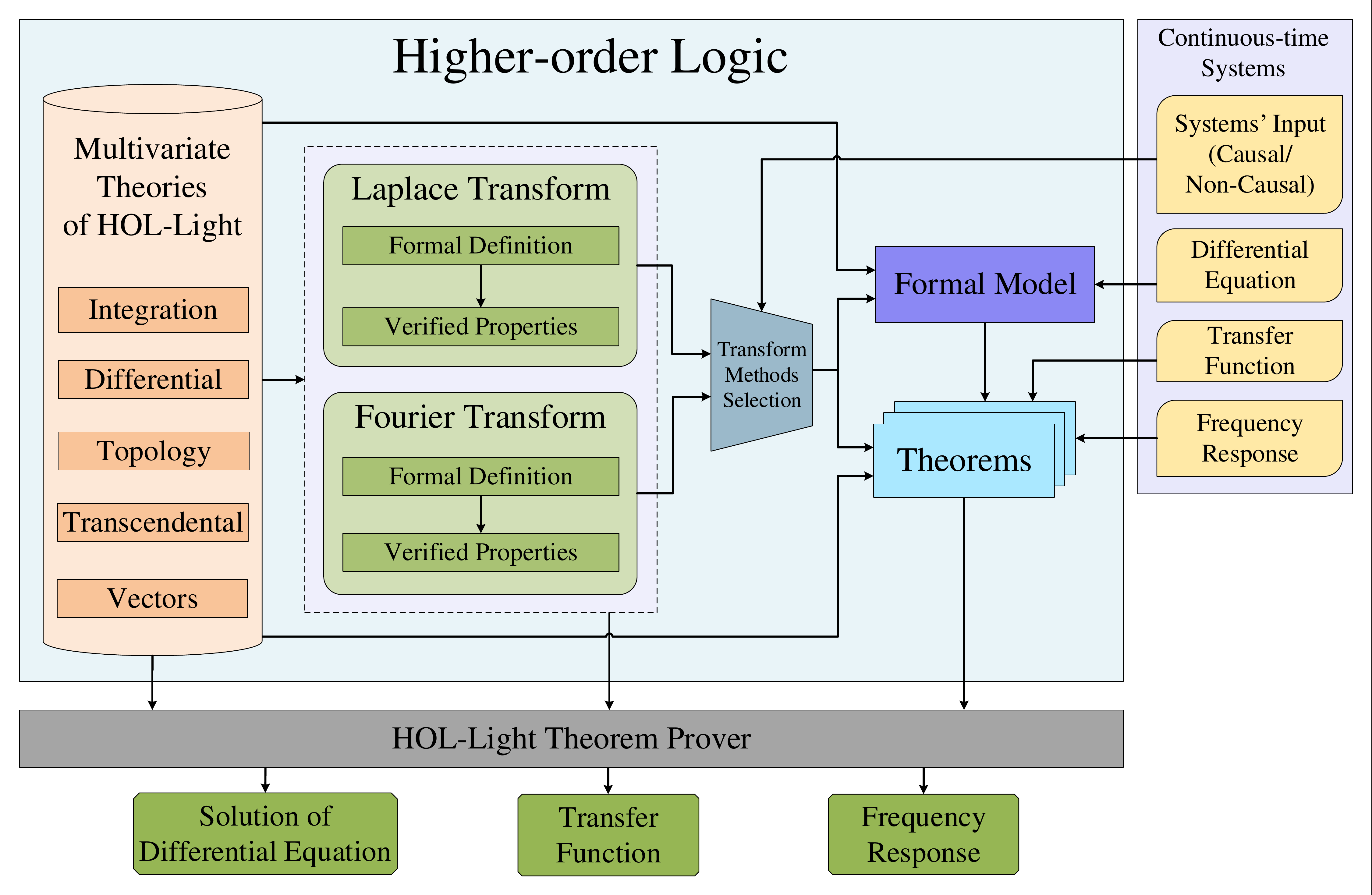}}
\caption{Transform Methods based Formal Analysis }
\label{FIG:proposed_framework}
\end{figure}

Fig.~\ref{FIG:proposed_framework} depicts the proposed approach for the transform methods based analysis of the continuous-time systems using the HOL Light theorem prover. The user provides the differential equation that models the dynamics of the system, which needs to be analyzed, and the corresponding input to the system. This differential equation is modeled in higher-order logic using the multivariate calculus theories of HOL Light. In the next step, we need to verify the corresponding properties, such as transfer function, frequency response or the solution of the corresponding differential equations. Our formalization of the Laplace and Fourier transform methods is used to develop the formal reasoning related to this verification.
Our formal approach allows the user to perform the analysis of a continuous-time system by selecting the suitable transform method (Laplace or Fourier) depending on the type of the system's input, i.e., if the input to the system is a causal function, then the Laplace transform is used. Similarly, in the case of the non-causal input, Fourier transform can be used.

\section{Results and Discussions}\label{SEC:formal_defs_of_transform_methods}

In this section, we present the existing formal definitions and some of the formally verified classical properties of the Laplace and Fourier transforms.
We also give some suggestions that can improve these definitions in terms of reasoning effort required to verify their properties.

\subsection{Laplace Transform} \label{SUBSEC:def_laplace_tran}

Laplace transform of a function $f(t): \mathds{R}^1 \rightarrow \mathds{C}$ is mathematically expressed as the following equation~\cite{beerends2003fourier}:

\vspace{0.5mm}

\begin{equation}\label{EQ:laplace_transform}
\mathcal{L} [f (t)] = (\mathcal{L} f)(s) = F(s) = \int_{0}^{\infty} {f(t)e^{-s t}} dt, \ s \ \epsilon \ \mathds{C}
\end{equation}

\vspace{0.5mm}

\noindent where $s$ is a complex variable. The limit of integration is from $0$ to ${\infty}$. The above equation can alternatively represented as:

\vspace{0.5mm}

\begin{equation}\label{EQ:laplace_transform_alt}
F(s) = \lim_{b \rightarrow \infty} \int_{0}^{b} {f(t)e^{-s t}} dt
\end{equation}

\vspace{0.5mm}

We formalize Equation~\ref{EQ:laplace_transform} using its alternate representation (Equation~\ref{EQ:laplace_transform_alt}), as follows~\cite{taqdees2013formalization}:

\vspace{1.0mm}

\begin{definition}
\label{DEF:laplace_transform}
\emph{Laplace Transform} \\{\small
\textup{\texttt{$\vdash$ $\forall$ s f. laplace f s =  \\
$\mathtt{\ }$\hspace{0.5cm} lim at\_posinfinity ($\lambda$b. integral (interval [lift (\&0), lift b]) \\
$\mathtt{\ }$\hspace{5.0cm} ($\lambda$t. cexp (--(s $\ast$ Cx (drop t))) $\ast$ f t))
}}}
\end{definition}

\vspace{1.0mm}

\noindent In the above definition, \texttt{integral} represents the vector integral. It takes the integrand function $\texttt{f}$ : $\mathds{R}^N \rightarrow \mathds{R}^M$, and a vector-space $\mathtt{i}: \mathds{R}^N \rightarrow \mathds{B}$, which defines the region of convergence, and returns the integral of $\mathtt{f}$ on $\mathtt{i}$ as a vector $\mathds{R}^M$~\cite{harrison17integ}.
The function \texttt{lim} in Definition~\ref{DEF:laplace_transform} takes a vector function $\texttt{f} : \mathds{A} \rightarrow \mathds{R}^M$ and $\texttt{net}$ : $\mathds{A}$ and returns $\texttt{l}$ of data-type $\mathds{R}^M$, i.e., the value to which $\texttt{f}$ converges at the given $\texttt{net}$. The function $\mathtt{lift}$ accepts a variable of type $\mathds{R}$ and maps it to a 1-dimensional vector with the input variable as its single component. Similarly, $\mathtt{drop}$ takes a 1-dimensional vector and returns its single element as a real number~\cite{harrison17vector}.

The Laplace transform of a function $f$ exists, if the function $\mathtt{f}$ is piecewise smooth and of exponential order on the positive real line~\cite{beerends2003fourier}.
The existence of the Laplace transform is formally defined as follows~\cite{taqdees2013formalization}:

\vspace{1.0mm}

\begin{definition}
\label{DEF:laplace_existence}
\emph{Laplace Exists} \\{\small
\textup{\texttt{$\vdash$ $\forall$ s f. laplace\_exists f s $\Leftrightarrow$ \\
$\mathtt{\ }$\hspace{0.5cm} ($\forall$ b. f piecewise\_differentiable\_on interval [lift (\&0), lift b]) $\wedge$ \\
$\mathtt{\ }$\hspace{0.5cm} ($\exists$ M a. Re s > drop a $\wedge$ exp\_order f M a)
}}}
\end{definition}

\vspace{1.0mm}

The function $\texttt{exp\_order}$ in the above definition is formally defined as~\cite{taqdees2013formalization}:

\vspace{1.0mm}

\begin{definition}
\label{DEF:exp_order_condition}
\emph{Exponential Order Function} \\{\small
\textup{\texttt{$\vdash$ $\forall$ f M a. exp\_order f M a $\Leftrightarrow$ \&0 < M $\wedge$ \\
$\mathtt{\ }$\hspace{2.0cm} ($\forall$ t. \&0 <= t $\Rightarrow$ norm (f (lift t)) <= M $\ast$ exp (drop a $\ast$ t))
}}}
\end{definition}

\vspace{1.0mm}

We used Definitions~\ref{DEF:laplace_transform},~\ref{DEF:laplace_existence} and~\ref{DEF:exp_order_condition} to formally
verify some of the classical properties of the Laplace transform, given in Table~\ref{TAB:properties_of_Laplace_transform}, which mainly include the linearity, frequency shifting, differentiation and integration in the time domain.
The formalization of the Laplace transform took around 5000 lines of code and approximately 450 man-hours.

\vspace{3.5mm}


\begin{scriptsize}
    \begin{longtable}{|p{2cm}|p{3cm}|p{7cm}|p{3cm}|}
\caption{Properties of Laplace Transform} \vspace{0.2cm}
\label{TAB:properties_of_Laplace_transform}
\endfirsthead
\endhead
    \hline
    \hline
    \multicolumn{1}{l}{Mathematical Form}   &
    \multicolumn{1}{l}{\hspace{-0.1cm} Formalized Form}

     \\ \hline \hline



   \multicolumn{2}{c}{\textbf{Limit Existence of Integral of Laplace Transform}}  \\ \hline

   \multicolumn{1}{l}{ {$\begin{array} {lcl} \textit{$ \exists l. \left(\int_0^{\infty} {f(t) e^{- s t}} \rightarrow l\right) $ }
 \end{array}$}  }  &

   \multicolumn{1}{l}{{$\begin{array} {lcl} \textup{\texttt{$\vdash$ $\forall$ f s. laplace\_exists f s   }} \\
\textup{\texttt{$\mathtt{\ }$\hspace{-0.1cm} $\Rightarrow$    ($\exists$l. (($\lambda$b. integral (interval [lift (\&0),lift b]) \hspace{-0.2cm}  }}  \\
\textup{\texttt{$\mathtt{\ }$\hspace{0.0cm} ($\lambda$t. cexp (--(s $\ast$ Cx (drop t))) $\ast$ f t)) $\rightarrow$ l) at\_posinfinity)  }}
 \end{array}$}}    \\ \hline



\multicolumn{2}{c}{\textbf{Linearity}} \\ \hline

   \multicolumn{1}{l}{ {$\begin{array} {lcl} \textit{$ \mathcal{L} [ \alpha f(t) + \beta g(t)] = $ } \\
\textit{$\mathtt{\ }$\hspace{0.4cm} $\alpha F(s) + \beta G(s) $     }
 \end{array}$}  }  &

   \multicolumn{1}{l}{{ $\begin{array} {lcl} \textup{\texttt{\hspace{-0.1cm}$\vdash$ $\forall$ f g s a b. laplace\_exists f s $\wedge$ laplace\_exists g s     }} \\
\textup{\texttt{$\mathtt{\ }$\hspace{0.4cm} $\Rightarrow$  laplace ($\lambda$t. a $\ast$ f t + b $\ast$ g t) s =  }} \\
\textup{\texttt{$\mathtt{\ }$\hspace{1.0cm} a $\ast$ laplace f s + b $\ast$ laplace g s  }}
 \end{array}$}}    \\ \hline



\multicolumn{2}{c}{\textbf{Frequency Shifting}} \\ \hline

    \multicolumn{1}{l}{ {$\begin{array} {lcl} \textit{$ \mathcal{L} [ e^{s_0 t} f(t)] =  $     } \\
    \textit{$\mathtt{\ }$\hspace{0.7cm} $ F(s - s_0) $     }
 \end{array}$} }   &

   \multicolumn{1}{l}{{$\begin{array} {lcl} \textup{\texttt{$\vdash$ $\forall$ f s s0. laplace\_exists f s      }} \\
\textup{\texttt{$\mathtt{\ }$\hspace{0.2cm} $\Rightarrow$ laplace ($\lambda$t. cexp (s0 $\ast$ Cx (drop t)) $\ast$ f t) s =  }} \\
\textup{\texttt{$\mathtt{\ }$\hspace{0.6cm} laplace f (s - s0)  }}
 \end{array}$}}    \\ \hline



\multicolumn{2}{c}{\textbf{First-order Differentiation}} \\ \hline

    \multicolumn{1}{l}{    {$\begin{array} {lcl} \textit{$ \mathcal{L} \left[\dfrac{d}{dt}f(t) \right] =  $ } \\
        \textit{$\mathtt{\ }$\hspace{0.4cm} $ s F(s) - f(0) $     }
 \end{array}$}  }   &

    \multicolumn{1}{l}{{$\begin{array} {lcl} \textup{\texttt{$\vdash$ $\forall$ f s. laplace\_exists f s $\wedge$ ($\forall$t. f differentiable at t) $\wedge$  }} \\
\textup{\texttt{$\mathtt{\ }$\hspace{0.2cm} laplace\_exists ($\lambda$t. vector\_derivative f (at t)) s   }}  \\
\textup{\texttt{$\mathtt{\ }$\hspace{1.0cm} $\Rightarrow$ laplace ($\lambda$t. vector\_derivative f (at t)) s =  }}  \\
\textup{\texttt{$\mathtt{\ }$\hspace{3.7cm} s $\ast$ laplace f s - f (lift (\&0))  }}
 \end{array}$}}    \\ \hline



\multicolumn{2}{c}{\textbf{Higher-order Differentiation}} \\ \hline

    \multicolumn{1}{l}{{$\begin{array} {lcl}
    \mathcal{L} [\dfrac{d^n}{{dt}^n}f(t)] = s^n F(s) \\
  \hspace{0.2cm}  - \sum_{k = 1}^{n}{ s^{k - 1} \dfrac{d^{n - k} f (0)}{{dx}^{n - k}} }
     \end{array}$}}    &

    \multicolumn{1}{l}{{$\begin{array} {lcl} \textup{\texttt{$\vdash$ $\forall$ f s n. laplace\_exists\_higher\_deriv n f s $\wedge$  }} \\
 \textup{\texttt{$\mathtt{\ }$\hspace{0.1cm} ($\forall$t. higher\_derivative\_differentiable n f t)  }} \\
\textup{\texttt{$\mathtt{\ }$\hspace{0.1cm} $\Rightarrow$ laplace ($\lambda$t. higher\_vector\_derivative n f t) s =  }} \\
\textup{\texttt{$\mathtt{\ }$\hspace{0.3cm} s pow n $\ast$ laplace f s - vsum (1..n) ($\lambda$k. s pow (k - 1) $\ast$  }} \\
\textup{\texttt{$\mathtt{\ }$\hspace{2.0cm} higher\_vector\_derivative (n - k) f (lift (\&0))) }}
 \end{array}$}}    \\ \hline



\multicolumn{2}{c}{\textbf{Integration in Time Domain}} \\ \hline

    \multicolumn{1}{l}{$ \mathcal{L} \left[ \int_{0}^{t}{f (\tau) d\tau} \right] = \dfrac{1}{s} F(s)  $ }   &

    \multicolumn{1}{l}{{$\begin{array} {lcl} \textup{\texttt{$\vdash$ $\forall$ f s. \&0 < Re s $\wedge$ laplace\_exists f s $\wedge$   }}  \\
\textup{\texttt{$\mathtt{\ }$\hspace{0.2cm} laplace\_exists ($\lambda$x. integral (interval [lift (\&0),x]) f) s $\wedge$  }}  \\
\textup{\texttt{$\mathtt{\ }$\hspace{0.2cm} ($\forall$x. f continuous\_on interval [lift (\&0),x])  }}  \\
\textup{\texttt{$\mathtt{\ }$\hspace{0.0cm} $\Rightarrow$ laplace ($\lambda$x. integral (interval [lift (\&0),x]) f) s =  }}  \\
\textup{\texttt{$\mathtt{\ }$\hspace{3.8cm}  inv s $\ast$ laplace f s  }}
\end{array}$}}    \\ \hline


    \end{longtable}

\end{scriptsize}


The formal definition of the Laplace transform presented as Definition~\ref{DEF:laplace_transform} is modeled using the notion of the limit. However, the HOL Light definition of the integral function (\texttt{integral}) implicitly encompasses infinite limits of integration, so we do not require to include another limiting process in its definition. Moreover, the region of integration (the positive real line) given in Equation~\ref{EQ:laplace_transform} can be modeled using the notion of set. So the mathematical definition of the Laplace transform, given by Equation~\ref{EQ:laplace_transform}, can alternatively be modeled in HOL Light as:

\begin{flushleft}
{\small
\textup{\texttt{$\vdash$ $\forall$ s f. laplace\_transform f s = \\
$\mathtt{\ }$\hspace{0.7cm} integral \{t| \&0 <= drop t\} ($\lambda$t. cexp (--(s $\ast$ Cx (drop t))) $\ast$ f t) }}}
\end{flushleft}

In the above definition, the region of integration, i.e., $[0, \infty)$ is modeled as \texttt{\small{\{t | \&0 <= drop t\}}} and this definition is equivalent to Definition~\ref{DEF:laplace_transform}. Moreover, this revised definition considerably simplifies the reasoning process in the verification of the properties of the Laplace transform since it does not involve the notion of limit.

\subsection{Fourier Transform} \label{SUBSEC:def_fourier_tran}

The Fourier transform of a function $f(t): \mathds{R}^1 \rightarrow \mathds{C}$ is mathematically defined as:

\begin{equation}\label{EQ:fourier_transform}
\mathcal{F} [f (t)] = (\mathcal{F} f)(\omega) = F(\omega) = \int_{-\infty}^{+\infty} {f(t)e^{-i \omega t}} dt, \ \omega \ \epsilon \  \mathds{R}
\end{equation}

\noindent where $\omega$ is a real variable. The limit of integration is from ${-\infty}$ to ${+\infty}$. We formalize Equation (\ref{EQ:fourier_transform}) as the following HOL Light function~\cite{rashid2016formalization}:

\begin{definition}
\label{DEF:fourier_transform}
\emph{Fourier Transform} \\{\small
\textup{\texttt{$\vdash$ $\forall$ w f. fourier f w = \\
$\mathtt{\ }$\hspace{1.5cm} integral UNIV ($\lambda$t. cexp (--((ii $\ast$ Cx w) $\ast$ Cx (drop t))) $\ast$ f t)
}}}
\end{definition}

The Fourier transform of a function $ f $ exists, i.e., the integrand of Equation \ref{EQ:fourier_transform} is integrable, and the integral has some converging limit value, if $f$ is piecewise smooth and is absolutely integrable on the whole real line~\cite{beerends2003fourier,rashid2016formalization}.
The Fourier existence condition can thus be formalized in HOL Light as follows:

\begin{definition}
\label{DEF:fourier_exists}
\emph{Fourier Exists} \\{\small
\textup{\texttt{$\vdash$ $\forall$ f. fourier\_exists f = \\
$\mathtt{\ }$\hspace{0.2cm} ($\forall$ a b. f piecewise\_differentiable\_on interval [lift a, lift b]) $\wedge$ \\
$\mathtt{\ }$\hspace{1.55cm} f absolutely\_integrable\_on \{x | \&0 <= drop x\} $\wedge$ \\
$\mathtt{\ }$\hspace{1.55cm} f absolutely\_integrable\_on \{x | drop x <= \&0\}
}}}
\end{definition}

\noindent In the above function, the first conjunct expresses the piecewise smoothness condition for the function $\texttt{f}$.
Whereas, the next two conjuncts represent the condition that the function $\texttt{f}$ is absolutely integrable on the whole real line.

We used Definitions~\ref{DEF:fourier_transform} and~\ref{DEF:fourier_exists} to verify some of the classical properties of Fourier transform, given in Table~\ref{TAB:properties_of_Fourier_transform}, such as existence, linearity, frequency shifting, modulation, time reversal and differentiation in time-domain. The formalization took around 3000 lines of code and approximately 250 man-hours.


\begin{scriptsize}
    \begin{longtable}{|p{2.2cm}|p{3.2cm}|p{7.2cm}|p{3.2cm}|}
\caption{Properties of Fourier Transform} \vspace{0.2cm}
\label{TAB:properties_of_Fourier_transform}
\endfirsthead
\endhead
    \hline
    \hline
    \multicolumn{1}{l}{Mathematical Form}   &
    \multicolumn{1}{l}{Formalized Form}

     \\ \hline \hline



   \multicolumn{2}{c}{\textbf{Integrability}}  \\ \hline

   \multicolumn{1}{l}{ {$\begin{array} {lcl} \textit{$f(t) e^{- i \omega t}\ integrable\  $ } \\
\textit{$\mathtt{\ }$\hspace{0.4cm} $ on\ (-\infty, \infty) $     }
 \end{array}$}  }  &

   \multicolumn{1}{l}{{$\begin{array} {lcl} \textup{\texttt{$\vdash$ $\forall$ f w. fourier\_exists f s     }} \\
\textup{\texttt{$\mathtt{\ }$\hspace{0.2cm} $\Rightarrow$ ($\lambda$t. cexp (--((ii $\ast$ Cx w) $\ast$ Cx (drop t))) $\ast$ f t) }}  \\
\textup{\texttt{$\mathtt{\ }$\hspace{2.0cm} integrable\_on UNIV \hspace{-0.2cm}  }}
 \end{array}$}}    \\ \hline



\multicolumn{2}{c}{\textbf{Linearity}} \\ \hline

   \multicolumn{1}{l}{ {$\begin{array} {lcl} \textit{$ \mathcal{F} [ \alpha f(t) + \beta g(t)] = $ } \\
\textit{$\mathtt{\ }$\hspace{0.4cm} $\alpha F(\omega) + \beta G(\omega) $     }
 \end{array}$}  }  &

   \multicolumn{1}{l}{{ $\begin{array} {lcl} \textup{\texttt{$\vdash$ $\forall$ f g w a b. fourier\_exists f $\wedge$ fourier\_exists g     }} \\
\textup{\texttt{$\mathtt{\ }$\hspace{0.4cm} $\Rightarrow$  fourier ($\lambda$t. a $\ast$ f t + b $\ast$ g t) w =  }} \\
\textup{\texttt{$\mathtt{\ }$\hspace{2.2cm} a $\ast$ fourier f w + b $\ast$ fourier g w  }}
 \end{array}$}}    \\ \hline



\multicolumn{2}{c}{\textbf{Frequency Shifting}} \\ \hline

    \multicolumn{1}{l}{ {$\begin{array} {lcl} \textit{$ \mathcal{F} [ e^{i \omega _0 t} f(t)] = F(\omega - \omega _0) $     }
 \end{array}$} }   &

   \multicolumn{1}{l}{ {$\begin{array} {lcl} \textup{\texttt{$\vdash$ $\forall$ f w w0. fourier\_exists f $\Rightarrow$     }} \\
\textup{\texttt{$\mathtt{\ }$\hspace{0.2cm} fourier ($\lambda$t. cexp ((ii $\ast$ Cx w0) $\ast$ Cx (drop t)) $\ast$ f t) w = \hspace{-1.0cm} }} \\
\textup{\texttt{$\mathtt{\ }$\hspace{0.2cm} fourier f (w - w0)  }}
 \end{array}$} }    \\ \hline



\multicolumn{2}{c}{\textbf{Modulation (Cosine and Sine Based Modulation)}} \\ \hline

    \multicolumn{1}{l}{    {$\begin{array} {lcl}  \textit{$ \mathcal{F}[cos(\omega_0 t) f(t)] = $ } \\
\textit{$\mathtt{\ }$\hspace{0.4cm} $\dfrac{F(\omega - \omega _0) + F(\omega + \omega _0)}{2} $     }
 \end{array}$}  }   &

    \multicolumn{1}{l}{{$\begin{array} {lcl} \textup{\texttt{$\vdash$ $\forall$ f w w0. fourier\_exists f     }} \\
\textup{\texttt{$\mathtt{\ }$\hspace{0.5cm} $\Rightarrow$  fourier ($\lambda$t. ccos (Cx w0 $\ast$ Cx (drop t)) $\ast$ f t) w =  }} \\
\textup{\texttt{$\mathtt{\ }$\hspace{2.5cm} $\mathtt{\dfrac{fourier\ f\ (w - w0)\ +\ fourier\ f\ (w + w0)}{Cx\ (\&2)}}$  }}
 \end{array}$}}    \\ \hline


    \multicolumn{1}{l}{    {$\begin{array} {lcl}  \textit{$ \mathcal{F}[sin(\omega_0 t) f(t)] = $ } \\
\textit{$\mathtt{\ }$\hspace{0.4cm} $ \dfrac{F(\omega - \omega _0) - F(\omega + \omega _0)}{2i} $     }
 \end{array}$}  }   &

    \multicolumn{1}{l}{{$\begin{array} {lcl} \textup{\texttt{$\vdash$ $\forall$ f w w0. fourier\_exists f     }} \\
\textup{\texttt{$\mathtt{\ }$\hspace{0.5cm} $\Rightarrow$  fourier ($\lambda$t. csin (Cx w0 $\ast$ Cx (drop t)) $\ast$ f t) w =  }} \\
\textup{\texttt{$\mathtt{\ }$\hspace{2.5cm} $\mathtt{\dfrac{fourier\ f\ (w - w0)\ -\ fourier\ f\ (w + w0)}{Cx\ (\&2) \ast ii}}$  }}

 \end{array}$}}    \\ \hline



\multicolumn{2}{c}{\textbf{Time Reversal}} \\ \hline

    \multicolumn{1}{l}{    $ \mathcal{F}[f(-t)] = F(-\omega) $      }   &

    \multicolumn{1}{l}{{$\begin{array} {lcl} \textup{\texttt{$\vdash$ $\forall$ f w. fourier\_exists f   }} \\
\textup{\texttt{$\mathtt{\ }$\hspace{0.5cm} $\Rightarrow$   fourier ($\lambda$t. f (--t)) w = fourier f (--w) \hspace{-1.0cm}   }}
 \end{array}$}}    \\ \hline



\multicolumn{2}{c}{\textbf{First-order Differentiation}} \\ \hline

    \multicolumn{1}{l}{ $ \mathcal{F} [\dfrac{d}{dt}f(t) ] = i \omega F(\omega) $  }   &

    \multicolumn{1}{l}{{$\begin{array} {lcl} \textup{\texttt{$\vdash$ $\forall$ f w. fourier\_exists f $\wedge$  }} \\
\textup{\texttt{$\mathtt{\ }$\hspace{0.2cm} fourier\_exists ($\lambda$t. vector\_derivative f (at t)) $\wedge$  }}  \\
\textup{\texttt{$\mathtt{\ }$\hspace{0.2cm} ($\forall$t. f differentiable at t) $\wedge$ }}  \\
\textup{\texttt{$\mathtt{\ }$\hspace{0.2cm} (($\lambda$t. f (lift t)) $\rightarrow$ vec 0) at\_posinfinity $\wedge$  }}  \\
\textup{\texttt{$\mathtt{\ }$\hspace{0.2cm} (($\lambda$t. f (lift t)) $\rightarrow$ vec 0) at\_neginfinity  }}  \\
\textup{\texttt{$\mathtt{\ }$\hspace{0.8cm} $\Rightarrow$ fourier ($\lambda$t. vector\_derivative f (at t)) w = \hspace{-1.0cm} }}  \\
\textup{\texttt{$\mathtt{\ }$\hspace{4.5cm} ii $\ast$ Cx w $\ast$ fourier f w  }}
 \end{array}$}}    \\ \hline



\multicolumn{2}{c}{\textbf{Higher-order Differentiation}} \\ \hline

    \multicolumn{1}{l}{$ \mathcal{F} [\dfrac{d^n}{{dt}^n}f(t)] = (i \omega)^n F(\omega)$ }   &

    \multicolumn{1}{l}{{$\begin{array} {lcl} \textup{\texttt{$\vdash$ $\forall$ f w n. fourier\_exists\_higher\_deriv n f $\wedge$  }} \\
\textup{\texttt{$\mathtt{\ }$\hspace{-0.2cm} ($\forall$t. differentiable\_higher\_derivative n f t) $\wedge$   }}  \\
\textup{\texttt{$\mathtt{\ }$\hspace{-0.2cm} ($\forall$p. p < n $\Rightarrow$ }}  \\
\textup{\texttt{$\mathtt{\ }$\hspace{0.0cm} (($\lambda$t. higher\_vector\_derivative p f (lift t)) $\rightarrow$ vec 0)  }}  \\
\textup{\texttt{$\mathtt{\ }$\hspace{1.2cm} at\_posinfinity) $\wedge$  }}  \\
\textup{\texttt{$\mathtt{\ }$\hspace{-0.2cm} ($\forall$p. p < n $\Rightarrow$  }}  \\
\textup{\texttt{$\mathtt{\ }$\hspace{0.5cm} (($\lambda$t. higher\_vector\_derivative p f (lift t)) $\rightarrow$ vec 0)  }} \\
\textup{\texttt{$\mathtt{\ }$\hspace{1.2cm} at\_neginfinity)  }} \\
\textup{\texttt{$\mathtt{\ }$\hspace{0.5cm} $\Rightarrow$ fourier ($\lambda$t. higher\_vector\_derivative n f t) w =   \hspace{-1.0cm} }} \\
\textup{\texttt{$\mathtt{\ }$\hspace{3.5cm} (ii $\ast$ Cx w) pow n $\ast$ fourier f w  }}
 \end{array}$}}    \\ \hline


    \end{longtable}

\end{scriptsize}


The absolute integrability condition in Definition~\ref{DEF:fourier_exists} is modeled using two conjuncts, i.e., the absolute integrability on the positive and negative real line, respectively. This condition can alternatively be modeled as:

\begin{flushleft}
{\small
\textup{\texttt{
$\mathtt{\ }$\hspace{0.00cm} f absolutely\_integrable\_on UNIV \\
}}}
\end{flushleft}

\noindent The function \texttt{UNIV} in the above condition presents the whole real line and is a composite modeling of the positive and negative real lines. Thus, this revised condition can better model the integrability condition and its equivalence to the earlier condition can be easily verified using some properties of the integrals.

\subsection{Relationship between Laplace and Fourier Transforms}\label{SEC:rel_laplace_fourier}

By restricting the complex-valued function $f(t): \mathds{R}^1 \rightarrow \mathds{C}$ and the Laplace variable $s:\mathds{R}^2$, we can find a very important relationship between Laplace and Fourier transforms.
If the function $f$ is causal, i.e., $f (t) = 0$ for all $t < 0$ and the real part of the Laplace variable $ \mathtt{s: R^2} $ is zero, i.e., $ \textit{Re s = 0} $, then the Laplace transform of function $f$ is equal to Fourier transform~\cite{thomas2016analysis}:

\begin{flushleft}
{
$ {(\mathcal{L} f)(s)\mid_{\textit{Re s = 0}} \ = \ (\mathcal{F} f)(Im \ s)} $
}
\end{flushleft}

The above relationship can be verified in HOL Light as follow:

\begin{flushleft}
{\small
\textup{\texttt{$\vdash$ $\forall$ f s. laplace\_exists f s $\wedge$ \\
$\mathtt{\ }$\hspace{0.2cm} ($\forall$t. t IN \{t | drop t <= \&0\} $\Rightarrow$ f t = vec 0) $\wedge$ ($\forall$t. Re s = \&0) \\
$\mathtt{\ }$\hspace{3.0cm} $\Rightarrow$ laplace\_transform f s = fourier\_transform f (Im s)
}}}
\end{flushleft}

\noindent This relationship is very crucial in a sense, if the function is causal, then the Laplace transform can be used in the analysis, rather than the Fourier transform.

\section{Achieved Goals, Current Status and Future Plans} \label{SEC:discussion}

The project started with the formalization of the Laplace transform and one of the major challenge faced during its formalization was that we were not very familiar with multivariable calculus theories of HOL Light and thus reasoning about theorems involving integration, differentiation and limits of the real and vector functions was very tedious for us as novice users of the system.
Moreover, we found that many basic properties required to reason about transform methods were not available in the multivariable theories in HOL Light and thus we ended up verifying many classical properties related to integration, differentiation and limit, including \textit{Comparison test for improper integrals}, \textit{Integration by substitution}, \textit{Integration by parts} and the \textit{Relationship between derivative of a real and vector functions}~\cite{taqdees2013formalization}.
The other major difficulty faced during these formalizations was the unavailability of detailed proofs for the properties of transform methods in literature. The available paper-and-pencil based proofs were found to be very abstract and we had to build the formal reasoning, at our own, for their verification.
Moreover, some of the assumptions of the properties of the Fourier transform were not explicitly mentioned in the literature, which we have extracted during the verification of these properties.
The foundational formalization of the Laplace and Fourier transform took about $8000$ lines of HOL Light code and $700$ man hours. The main benefit of this formalization was found in the ability to conduct formal transform method based analysis of many systems, including linear transfer converter~\cite{taqdees2013formalization}, which is widely used component in power electronics, the first and second-order Sallen-Key low-pass filters~\cite{taqdees2017tflac} and the automobile suspension system~\cite{rashid2016formalization}.
The foundational formalization of Laplace and Fourier transforms was found to be quite useful in this context and the analysis of these applications was found to be very straightforward and took only $1600$ lines of HOL Light code and $8$ man hours only.

The distinguishing feature of the transform methods based formal analysis, compared to traditional analysis methods, is the generic nature of the formally verified theorems. All the variables and functions are universally quantified and thus can be specialized to obtain the results for any given values.
Moreover, all of the required assumptions are guaranteed to be explicitly mentioned along with a formally verified theorem due to the inherent soundness of the theorem proving approach.
Moreover, the high expressiveness of the higher-order logic enables us to model the differential equation and the corresponding transfer function and frequency response in their true continuous form, whereas, in model checking, they are mostly discretized and modeled using a state-transition system, which may compromise the completeness of the analysis. A comparison of different analysis techniques for the transform methods is presented in Table~\ref{TAB:comparisons_of_analysis_techniques_transform_methods}. The evaluation of these techniques is performed based on various parameters, such as expressiveness, accuracy and automation.
For example, in model checking, we cannot truly model the integration and differentiation, and their discretization results into an abstracted model, which makes it less expressive. Moreover, in theorem proving, the verification is done interactively due to the undecidable nature of higher-order logic. We are mainly working on facilitating the user in this interactive verification part by providing formal reasoning support for Laplace and Fourier transforms.


\begin{scriptsize}
    \begin{longtable}{|p{2cm}|p{3cm}|p{7cm}|p{3cm}|p{3cm}|p{3cm}|}
\caption{Comparison of Analysis Techniques for Transform Methods} \vspace{0.2cm}
\label{TAB:comparisons_of_analysis_techniques_transform_methods}
\endfirsthead
\endhead
    \hline
    \hline
    \multicolumn{1}{l}{ }   &
    \multicolumn{1}{l}{{$\begin{array} {lcl} \textrm{ \hspace{0.0cm} Paper-and-Pencil } \\
\textrm{ \hspace{0.0cm} Proof }
 \end{array}$}} &
    \multicolumn{1}{l}{\hspace{0.0cm} Simulation} &
    \multicolumn{1}{l}{{$\begin{array} {lcl} \textrm{ \hspace{0.0cm} Computer Algebra } \\
 \textrm{ \hspace{0.0cm} System }
 \end{array}$}} &
    \multicolumn{1}{l}{{$\begin{array} {lcl} \textrm{ \hspace{0.0cm} Model } \\
\textrm{ \hspace{0.0cm} Checking }
 \end{array}$}} &
    \multicolumn{1}{l}{{$\begin{array} {lcl} \textrm{ \hspace{0.0cm} Theorem } \\
\textrm{ \hspace{0.0cm} Proving }
 \end{array}$}}

     \\ \hline \hline



   \multicolumn{1}{l}{  Expressiveness }  &
   \multicolumn{1}{c}{ \hspace{0.0cm} \begin{normalsize} \cmark \end{normalsize} }  &
    \multicolumn{1}{c}{ \hspace{0.0cm} \begin{normalsize} \cmark \end{normalsize} } &
    \multicolumn{1}{c}{ \hspace{0.0cm} \begin{normalsize} \cmark \end{normalsize} } &
       \multicolumn{1}{c}{ \hspace{0.0cm} \begin{normalsize}   \end{normalsize} } &
      \multicolumn{1}{c}{ \hspace{0.0cm} \begin{normalsize} \cmark \end{normalsize} }
       \\ \hline



   \multicolumn{1}{l}{ Accuracy }  &
   \multicolumn{1}{c}{ \hspace{0.0cm} \begin{normalsize} \cmark (\textbf{?}) \end{normalsize} }  &
    \multicolumn{1}{c}{ \hspace{0.0cm} \begin{normalsize}   \end{normalsize} } &
    \multicolumn{1}{c}{ \hspace{0.0cm} \begin{normalsize}   \end{normalsize} } &
       \multicolumn{1}{c}{ \hspace{0.0cm} \begin{normalsize} \cmark \end{normalsize} } &
       \multicolumn{1}{c}{ \hspace{0.0cm} \begin{normalsize} \cmark \end{normalsize} }
      \\ \hline



   \multicolumn{1}{l}{ Automation }  &
   \multicolumn{1}{c}{ \hspace{0.0cm} \begin{normalsize}   \end{normalsize} }  &
    \multicolumn{1}{c}{ \hspace{0.0cm} \begin{normalsize} \cmark \end{normalsize} } &
    \multicolumn{1}{c}{ \hspace{0.0cm} \begin{normalsize} \cmark \end{normalsize} } &
       \multicolumn{1}{c}{ \hspace{0.0cm} \begin{normalsize} \cmark \end{normalsize} } &
       \multicolumn{1}{c}{ \hspace{0.0cm} \begin{normalsize}  \end{normalsize} }
      \\ \hline


    \end{longtable}

\end{scriptsize}


\noindent We are currently focussing on the following three tasks:
\begin{itemize}
  \item Formalization of inverse Laplace and Fourier transforms: We have formally verified the uniqueness property of the Laplace transform and are working on verifying it for the Fourier Transform. These properties would enable us to verify the analytical solutions of linear differential equations.
  \item Automation of the transform methods based formal analysis: We are in the process of developing some tactics to automate the transform methods based formal analysis of the continuous-time systems. These tactics would only require the differential equation, modeling the dynamics of the systems, and expressions for the corresponding transfer functions and frequency responses and would automatically verify the relationships between them. This would allow non-experts in theorem proving to benefit from our formal approach for the analysis of the systems.
  \item Formalization of Vectorial Laplace transform: The current formalization of the Laplace transform can only be used for the formal analysis of the single-input single-output (SISO) control systems. We are working on extending the reasoning support for the normal Laplace transform to complex vectors. The resulting formalization would help us to formally verify the transfer function of the multiple-input multiple output (MIMO) control systems, which are modeled using the state space representations.
\end{itemize}

\noindent To further extend the scope of transform methods based formal analysis of systems, we plan to work on the following two tasks in the future:

\begin{itemize}
  \item Linking the formal library of the Laplace transform with the formalization of Z-transform~\cite{siddique2014formalization}: This linkage will enable us to perform the formal analysis of the hybrid (exhibiting continuous and discrete behaviour) systems.
  \item Formalization of two-dimensional Fourier transform: This requires the formalization of double integral and its properties, which, to the best of our knowledge, have not been formalized in the current multivariable calculus theories of HOL Light. The two-dimensional Fourier transform would build upon this theory of double integration. This formalization will enable us to perform the formal analysis of many electromagnetic (e.g.,~\cite{bracewell1965fourier,jin2011theory}) and the optical systems (e.g.,~\cite{born1980principles,bracewell1965fourier}). Moreover, this formalization of double integral can also be used for the formal analysis of some other applications in physics, such as, quantum~\cite{gorini2012fundamental} and mechanics~\cite{pytel2016engineering}.
\end{itemize}

\section{Impact} \label{SEC:case_studies}

Our foundational formalization of the Laplace~\cite{taqdees2013formalization} and the Fourier~\cite{rashid2016formalization} transforms has been used for the formal analysis of the various systems and some of them are presented in Table~\ref{TAB:appl_trans_meth}.
Due to the availability of the higher-order-logic


\begin{scriptsize}
    \begin{longtable}{|p{2cm}|p{3cm}|p{7cm}|p{3cm}|}
\caption{Applications of Transform Methods}
\vspace*{0.2cm}
\label{TAB:appl_trans_meth}
\endfirsthead
\endhead
    \hline
    \hline



\multicolumn{2}{c}{\textbf{Linear Transfer Converter~\cite{taqdees2013formalization}}} \\ \hline

    \multicolumn{1}{l}{ $\begin{array} {lcl}  \begin{minipage}{.31\textwidth}
 \vspace{0.0cm}  \hspace*{-0.4cm}  \includegraphics[width=1.50\textwidth, height=32mm, trim={5 5 5 5},clip]{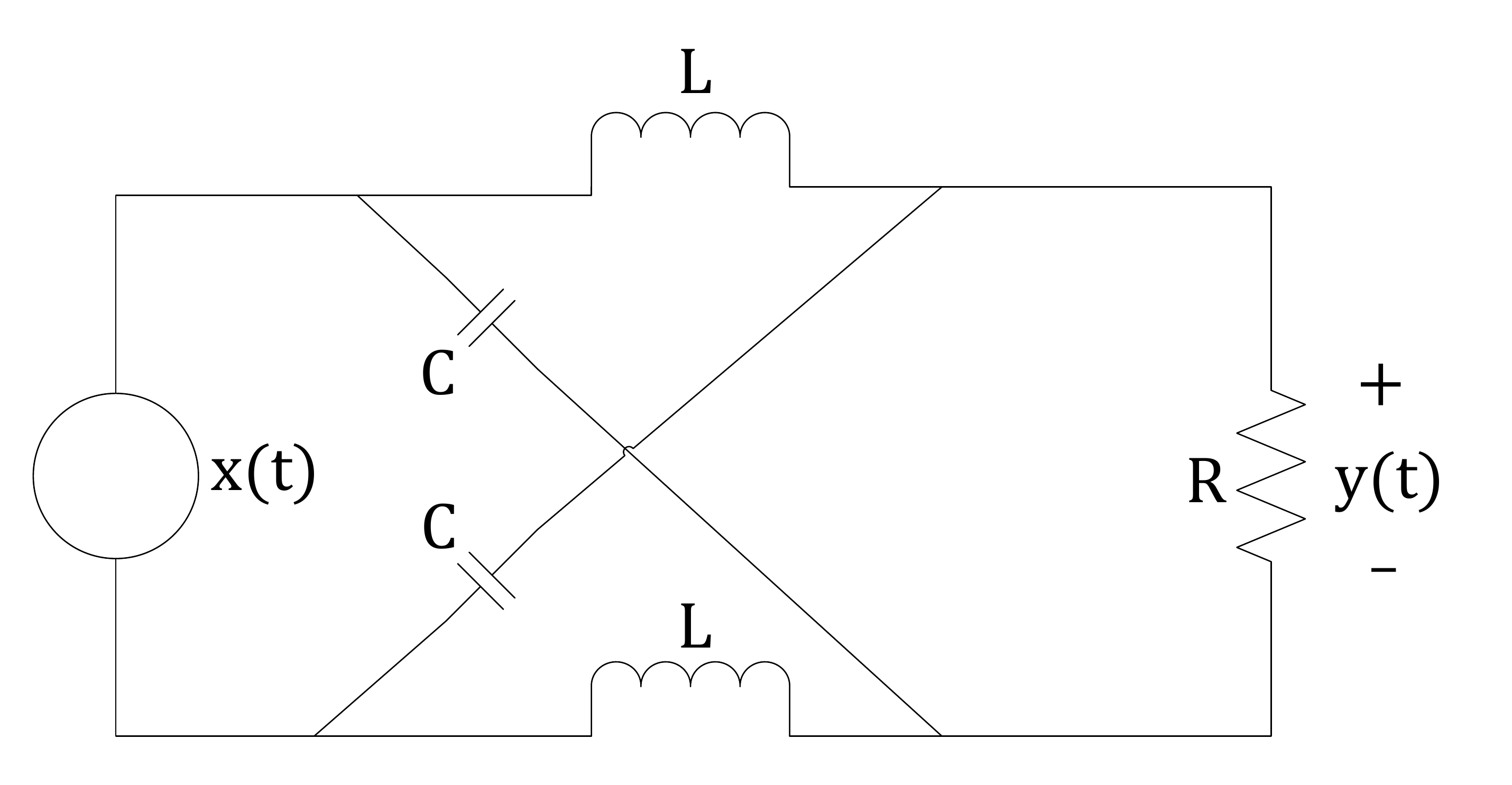}
    \end{minipage} \hspace{0.0cm} \\ \textup{\textrm{\hspace{0.0cm}Lines of code: 650    }} \\
\textup{\textrm{\hspace{-0.1cm} Man-hours: 2 }}
\vspace{0.0cm}  \end{array}$  }   &

   \multicolumn{1}{l}{{$\begin{array} {lcl} \textup{\texttt{ \hspace{1.30cm}$\vdash$ $\forall$ y u s R L C.    }} \\
\textup{\texttt{$\mathtt{\ }$\hspace{1.30cm} (\&0 < R) $\wedge$ (\&0 < L) $\wedge$  (\&0 < C) $\wedge$ }} \\
\textup{\texttt{$\mathtt{\ }$\hspace{1.30cm}  zero\_initial\_conditions 1 u $\wedge$ }} \\
\textup{\texttt{$\mathtt{\ }$\hspace{1.30cm}  zero\_initial\_conditions 1 y $\wedge$ }} \\
\textup{\texttt{$\mathtt{\ }$\hspace{1.30cm} (higher\_derivative\_laplace\_exists 2 u s) $\wedge$  }} \\
\textup{\texttt{$\mathtt{\ }$\hspace{1.30cm} (higher\_derivative\_laplace\_exists 2 y s) $\wedge$  }} \\
\textup{\texttt{$\mathtt{\ }$ \hspace{1.30cm} ($\forall$t. higher\_derivative\_differentiable 2 u t) $\wedge$  }} \\
\textup{\texttt{$\mathtt{\ }$ \hspace{1.30cm} ($\forall$t. higher\_derivative\_differentiable 2 y t) $\wedge$  }} \\
\textup{\texttt{$\mathtt{\ }$ \hspace{1.30cm} ($\forall$t. diff\_eq\_LTC y u L C R) $\wedge$  }} \\
\textup{\texttt{$\mathtt{\ }$ \hspace{1.30cm}  (non\_zero\_denominator u s R L C) }}  \vspace{0.1cm} \\
\textup{\texttt{$\mathtt{\ }$ \hspace{1.30cm} $\Rightarrow$ $\mathtt{\dfrac{\texttt{laplace y s}}{\texttt{laplace x s}} }$  =  }} \\
\textup{\texttt{$\mathtt{\ }$\hspace{2.0cm} $\mathtt{\dfrac{s\ pow\ 2 - Cx \Bigg( \dfrac{\&1}{L \ast C} \Bigg)}{Cx \Bigg( \dfrac{\&1}{L \ast C} \Bigg) - Cx \Bigg( \dfrac{\&2}{R \ast C} \Bigg) \ast s + s\ pow \ 2   } }$   }}
\vspace*{0.2cm}
 \end{array}$}}    \\ \hline



\multicolumn{2}{c}{\textbf{Automobile Suspension System~\cite{rashid2016formalization}}} \\ \hline

    \multicolumn{1}{l}{ $\begin{array} {lcl}  \begin{minipage}{.31\textwidth}
 \vspace{0.0cm}  \hspace*{-0.2cm}  \includegraphics[width=1.40\textwidth, height=38mm, trim={10 10 10 10},clip]{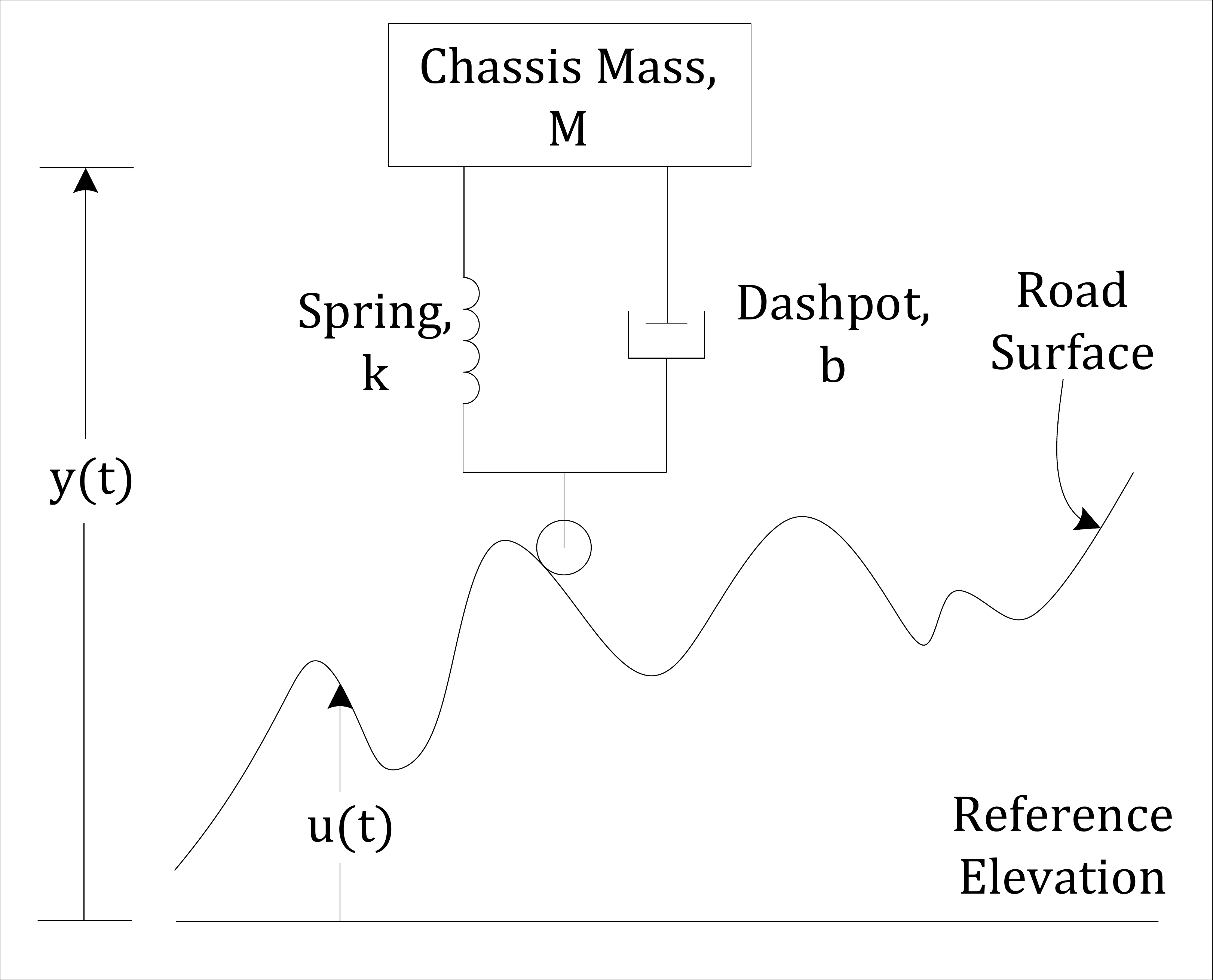}
    \end{minipage} \hspace{0.0cm} \\ \textup{\textrm{\hspace{0.0cm}Lines of code: 500    }} \\
\textup{\textrm{\hspace{-0.1cm} Man-hours: 2 }}
\vspace{0.0cm}  \end{array}$  }   &

   \multicolumn{1}{l}{{$\begin{array} {lcl} \textup{\texttt{\hspace{1.30cm}$\vdash$ $\forall$ y u w a.     }} \\
\textup{\texttt{$\mathtt{\ }$\hspace{1.30cm} (\&0 < M) $\wedge$ (\&0 < b) $\wedge$ (\&0 < k) $\wedge$   }} \\
\textup{\texttt{$\mathtt{\ }$\hspace{1.30cm} ($\forall$t. differentiable\_higher\_derivative 2 y t) $\wedge$  }} \\
\textup{\texttt{$\mathtt{\ }$\hspace{1.30cm} ($\forall$t. differentiable\_higher\_derivative 1 u t) $\wedge$  }} \\
\textup{\texttt{$\mathtt{\ }$\hspace{1.30cm} (fourier\_exists\_higher\_deriv 2 y) $\wedge$  }} \\
\textup{\texttt{$\mathtt{\ }$\hspace{1.30cm} (fourier\_exists\_higher\_deriv 1 u) $\wedge$  }} \\
\textup{\texttt{$\mathtt{\ }$ \hspace{1.30cm} ($\forall$p. p < 2 $\Rightarrow$ (($\lambda$t. higher\_vector\_derivative   }} \\
\textup{\texttt{$\mathtt{\ }$ \hspace{1.8cm} p y (lift t)) $\rightarrow$ vec 0) at\_posinfinity)) $\wedge$   }} \\
\textup{\texttt{$\mathtt{\ }$ \hspace{1.30cm} ($\forall$p. p < 2 $\Rightarrow$ (($\lambda$t. higher\_vector\_derivative   }} \\
\textup{\texttt{$\mathtt{\ }$ \hspace{1.8cm} p y (lift t)) $\rightarrow$ vec 0) at\_neginfinity)) $\wedge$ }} \\
\textup{\texttt{$\mathtt{\ }$\hspace{1.30cm} (($\lambda$t. u (lift t)) $\rightarrow$ vec 0) at\_posinfinity) $\wedge$  }}  \\
\textup{\texttt{$\mathtt{\ }$\hspace{1.30cm} (($\lambda$t. u (lift t)) $\rightarrow$ vec 0) at\_neginfinity) $\wedge$ }}  \\
\textup{\texttt{$\mathtt{\ }$\hspace{1.5cm} ($\forall$t. diff\_eq\_ASS y u b M k) $\wedge$  }} \\
\textup{\texttt{$\mathtt{\ }$\hspace{1.5cm}  (non\_zero\_denominator u w b M k) }} \vspace{0.1cm} \\
\textup{\texttt{$\mathtt{\ }$\hspace{1.30cm} $\Rightarrow$ $\mathtt{\dfrac{\texttt{fourier y w}}{\texttt{fourier x w}} }$  = }}   \vspace*{0.2cm}   \\
\textup{\texttt{$\mathtt{\ }$\hspace{1.5cm} $\mathtt{\dfrac{Cx \Bigg( \dfrac{b}{M} \Bigg) \ast ii \ast Cx\ w + Cx \Bigg( \dfrac{k}{M} \Bigg) }{Cx \Bigg( \dfrac{k}{M} \Bigg) + Cx \Bigg( \dfrac{b}{M} \Bigg) \ast ii \ast Cx\ w + (ii \ast Cx\ w)\ pow \ 2   } }$ }} \\
\vspace*{0.1cm}
 \end{array}$}}    \\ \hline



\multicolumn{2}{c}{\textbf{Second order Sallen-key Filter~\cite{taqdees2017tflac}}} \\ \hline

    \multicolumn{1}{l}{ $\begin{array} {lcl}  \begin{minipage}{.31\textwidth}
 \vspace{0.0cm}  \hspace*{-0.25cm}  \includegraphics[width=1.40\textwidth, height=27mm, trim={10 10 10 10},clip]{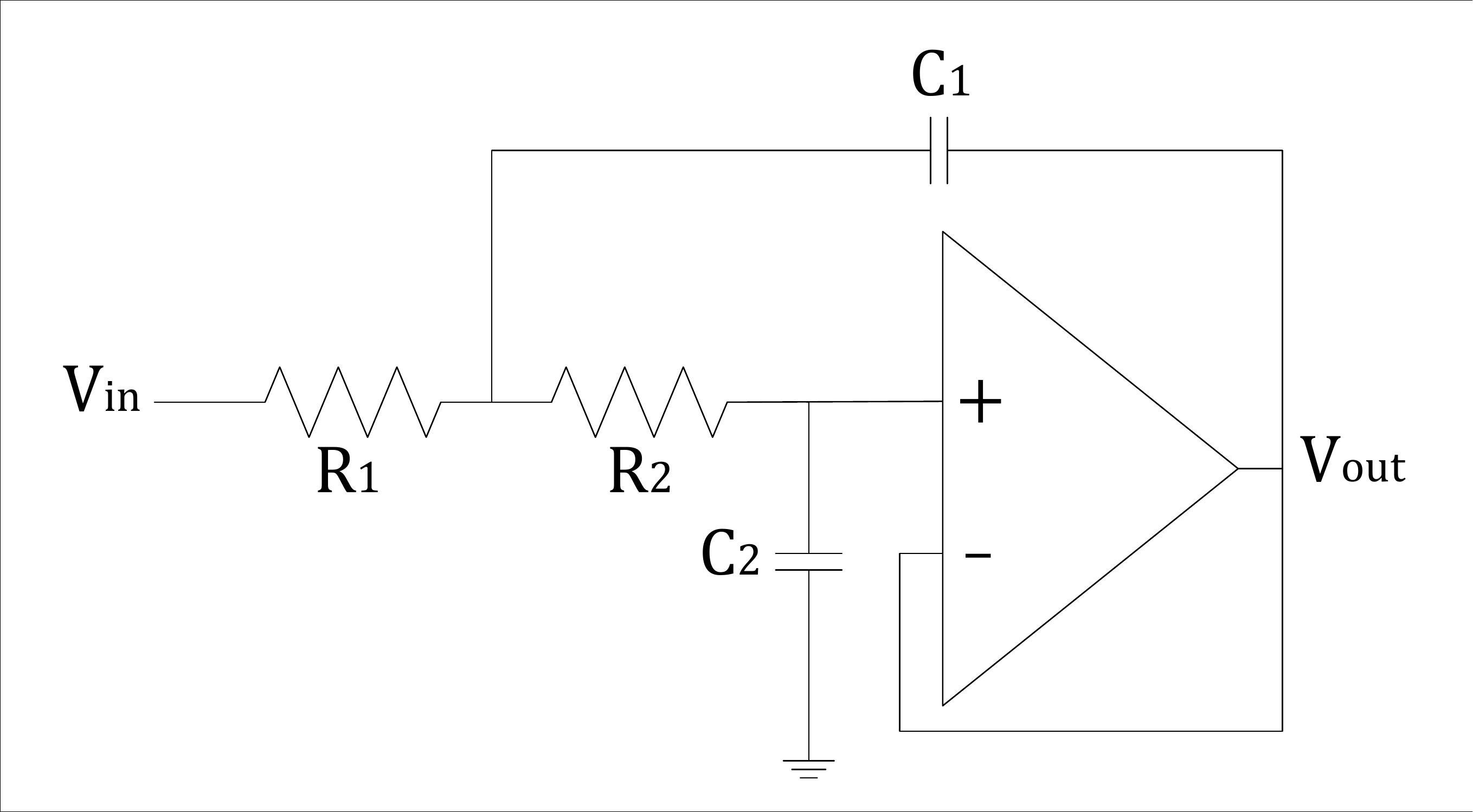}
    \end{minipage} \hspace{0.0cm} \\ \textup{\textrm{\hspace{0.0cm}Lines of code: 250    }} \\
\textup{\textrm{\hspace{-0.1cm} Man-hours: 2 }}
\vspace{0.0cm}  \end{array}$  }   &

   \multicolumn{1}{l}{{$\begin{array} {lcl} \textup{\texttt{\hspace{1.2cm}$\vdash$ $\forall$ R1 R2 C1 C2 Vin Vout s. (\&0 < R1) $\wedge$   }} \\
\textup{\texttt{$\mathtt{\ }$\hspace{1.4cm}  (\&0 < C1) $\wedge$ (\&0 < C1) $\wedge$ (\&0 < C2) $\wedge$  }} \\
\textup{\texttt{$\mathtt{\ }$\hspace{1.5cm}    zero\_initial\_conditions Vin Vout Va $\wedge$  }} \\
\textup{\texttt{$\mathtt{\ }$ \hspace{1.2cm} (laplace\_exists\_higher\_deriv 2 Vout s) $\wedge$ }} \\
\textup{\texttt{$\mathtt{\ }$ \hspace{1.2cm} (laplace\_exists\_higher\_deriv 2 Vin s) $\wedge$ }} \\
\textup{\texttt{$\mathtt{\ }$ \hspace{1.0cm} ($\forall$t. differentiable\_higher\_derivative 2 Vout t) $\wedge$  }} \\
\textup{\texttt{$\mathtt{\ }$ \hspace{1.0cm} ($\forall$t. differentiable\_higher\_derivative 2 Vin t) $\wedge$  }} \\
\textup{\texttt{$\mathtt{\ }$ \hspace{1.0cm} ($\forall$t. differentiable\_higher\_derivative 2 Va t) $\wedge$  }} \\
\textup{\texttt{$\mathtt{\ }$ \hspace{1.2cm}  (non\_zero\_denom Vin s R1 R2 C1 C2) $\wedge$  }} \\
\textup{\texttt{$\mathtt{\ }$ \hspace{1.2cm} ($\forall$t. SKF\_behav Vin Vout R1 R2 C1 C2)   }} \vspace{0.1cm} \\
\textup{\texttt{$\mathtt{\ }$\hspace{1.50cm} $\Rightarrow$ $\mathtt{\dfrac{\texttt{laplace Vout s}}{\texttt{laplace Vin s}} }$  =   }} \\
\textup{\texttt{$\mathtt{\ }$\hspace{3.0cm} $\mathtt{\dfrac{Cx (\&1)}{ \splitdfrac{ Cx \Bigg( R1 \ast C1 \ast R2 \ast C2 \Bigg) \ast s\ pow \ 2 + }{Cx \Bigg( C2 \ast (R1 + R2) \Bigg) \ast s + Cx (\&1) }  } }$  }}
\vspace*{0.1cm}
 \end{array}$}}    \\ \hline


    \end{longtable}

\end{scriptsize}


\noindent  formalization of transform methods, the analysis of these applications was very straightforward. It can be seen that these analyses took very few lines of code and very less manual effort, which clearly illustrates the effectiveness of our foundational formalization. These formalizations of the Laplace and Fourier transform can be further used for the analysis of the many other applications, including control systems, power electronics, signal processing and communication systems.

\section{Conclusion}\label{SEC:Conclusion}

This paper provides a synthetic presentation of our ongoing project on the formalization of transform methods using the HOL Light theorem prover.
We present the proposed approach for the transform methods based formal analysis of the continuous-time systems along with the foundational formal definitions of the Laplace and Fourier transform. The paper highlights the main objectives of the project that have been achieved so far, the challenges faced during this formalization, and the ongoing tasks and the future goals for this project. Once all the planned formalization tasks are accomplished, then these foundations can be used for the formal analysis of many safety-critical systems, such as control systems, power electronics, signal processing, electromagnetics and
optical systems.

\section*{Acknowledgements}

This work was supported by the National Research Program for Universities grant (number 1543) of Higher Education Commission (HEC), Pakistan.

\bibliographystyle{splncs03}
\bibliography{bibliotex}

\end{document}